\documentclass[11pt]{article}
\pdfoutput=1
\usepackage[centertags]{amsmath}
\usepackage[square, comma, sort&compress,numbers]{natbib}
\usepackage{array,multirow}
\numberwithin{equation}{section}
\usepackage{amssymb,amsfonts}
\usepackage{graphicx}
\usepackage{color}
\usepackage{mathtools}

\usepackage{epsfig}
\usepackage{bbold}

\usepackage{wrapfig}
\usepackage{float}
\usepackage{soul}
\usepackage{tkz-euclide}

\usepackage{tikz,pgf}
\usetikzlibrary{shapes}
\usetikzlibrary{calc}
\usetikzlibrary{decorations.pathmorphing}
\usetikzlibrary{decorations.pathreplacing,shapes.misc}
\usetikzlibrary{positioning}
\usetikzlibrary{arrows}
\usetikzlibrary{decorations.markings}
\usetikzlibrary{shadings}

\usetikzlibrary{intersections}

\def\be{\begin{equation}}
\def\ee{\end{equation}}
\def\ba{\begin{array}}
\def\ea{\end{array}}

\def\dps{\displaystyle}

\def\1{\tilde{1}}
\def\2{\tilde{2}}
\def\3{\tilde{3}}

\newdimen\tableauside\tableauside=1.0ex
\newdimen\tableaurule\tableaurule=0.4pt
\newdimen\tableaustep
\def\phantomhrule#1{\hbox{\vbox to0pt{\hrule height\tableaurule
width#1\vss}}}
\def\phantomvrule#1{\vbox{\hbox to0pt{\vrule width\tableaurule
height#1\hss}}}
\def\sqr{\vbox{%
  \phantomhrule\tableaustep

\hbox{\phantomvrule\tableaustep\kern\tableaustep\phantomvrule\tableaustep}%
  \hbox{\vbox{\phantomhrule\tableauside}\kern-\tableaurule}}}
\def\squares#1{\hbox{\count0=#1\noindent\loop\sqr
  \advance\count0 by-1 \ifnum\count0>0\repeat}}
\def\tableau#1{\vcenter{\offinterlineskip
  \tableaustep=\tableauside\advance\tableaustep by-\tableaurule
  \kern\normallineskip\hbox
    {\kern\normallineskip\vbox
      {\gettableau#1 0 }%
     \kern\normallineskip\kern\tableaurule}%
  \kern\normallineskip\kern\tableaurule}}
\def\gettableau#1 {\ifnum#1=0\let\next=\null\else
  \squares{#1}\let\next=\gettableau\fi\next}

\newcommand{\bref}[1]{\textbf{\ref{#1}}}

\def\cO{\mathcal{O}}

\numberwithin{equation}{section} \makeatletter
\@addtoreset{equation}{section}

\def\be{\begin{equation}}
\def\ee{\end{equation}}
\def\ba{\begin{array}}
\def\ea{\end{array}}
\def\dps{\displaystyle}

\def\ba{\begin{array}}
\def\ea{\end{array}}

\newmuskip\pFqmuskip
\newcommand*\pFq[6][8]{%
  \begingroup 
  \pFqmuskip=#1mu\relax
  \mathcode`\,=\string"8000
  \begingroup\lccode`\~=`\,
  \lowercase{\endgroup\let~}\pFqcomma
  {}_{#2}F_{#3}{\left[\genfrac..{0pt}{}{#4}{#5};#6\right]}%
  \endgroup
}
\newcommand{\pFqcomma}{\mskip\pFqmuskip}

\newmuskip\LpFqmuskip
\newcommand*\LpFq[6][8]{%
  \begingroup 
  \pFqmuskip=#1mu\relax
  \mathcode`\,=\string"8000
  \begingroup\lccode`\~=`\,
  \lowercase{\endgroup\let~}\pFqcomma
  {}_{}F_{}{\left[\genfrac..{0pt}{}{#4}{#5};#6\right]}%
  \endgroup
}

\newmuskip\Ftmuskip
\newcommand*\Ft[6][8]{%
  \begingroup 
  \pFqmuskip=#1mu\relax
  \mathcode`\,=\string"8000
  \begingroup\lccode`\~=`\,
  \lowercase{\endgroup\let~}\pFqcomma
  F_{2}{\left[\genfrac..{0pt}{}{#4}{#5};#6\right]}%
  \endgroup
}
\newmuskip\Ftmuskip
\newcommand*\FK[6][8]{%
  \begingroup 
  \pFqmuskip=#1mu\relax
  \mathcode`\,=\string"8000
  \begingroup\lccode`\~=`\,
  \lowercase{\endgroup\let~}\pFqcomma
  F_{K}{\left[\genfrac..{0pt}{}{#4}{#5};#6\right]}%
  \endgroup
}

\usepackage{jheppub}
\makeatletter
\def\@fpheader{\vspace{-.1cm}}
\makeatother

\title{Towards $\mathcal{W}_3$ classical blocks with semi-degenerate operators}

\author[a]{V. Belavin}
\author[b,c,d]{Mikhail Pavlov}

\affiliation[a]{Physics Department, Ariel University, 
Ariel 40700, Israel.}

\affiliation[b]{I.E. Tamm Department of Theoretical Physics, \\P.N. Lebedev Physical
Institute,\\ Leninsky ave. 53, 119991 Moscow, Russia}

\affiliation[c]{Institute for Theoretical and Mathematical Physics,\\
Lomonosov Moscow State University, \\
Leninskie Gory, GSP-1, 119991 Moscow, Russia}
\affiliation[d]{HSE University, 6 Usacheva str., Moscow 119048, Russia}

\emailAdd{vladimirbe@ariel.ac.il}
\emailAdd{pavlov@lpi.ru}

\abstract{We consider 4-point $\mathcal{W}_3$ classical blocks focusing on the blocks level-1 and level-2 semi-degenerate operators. We derive BPZ-type equations for the auxiliary 5-point blocks with one additional fully degenerate operator. The monodromy properties of these equations are encoded by the accessory parameters, related to the 4-point $\mathcal{W}_3$ classical blocks. We solve the BPZ-type equations via heavy-light perturbation theory and find the accessory parameters, which allows us to obtain the explicit expressions for the considered class of classical blocks.}

\makeatletter
\def\@fpheader{\vspace{-.1cm}}
\makeatother

\begin{document}

\maketitle 
\flushbottom

\flushbottom

\section{Introduction}

In CFT$_2$, the main observables are the correlation functions of local operators. These $n$-point correlation functions can be decomposed into conformal blocks \cite{Belavin:1984vu} (determined only by the symmetries of the theory). The $n$-point conformal blocks are basis functions in the space of correlation functions, while coefficients in the expansion are the $n$-point structure constants. These coefficients are constructed as a product of $(n-2)$ the $3$-point structure constants. Hence, in the bootstrap program \cite{Belavin:1984vu} the $3$-point structure constants and $n$-point conformal blocks are the fundamental objects.

In Virasoro CFT$_2$, the correlation functions/conformal blocks involving degenerate operators are subjected by Belavin-Polyakov-Zamolodchikov (BPZ) differential equations following from the null-vector conditions \cite{Belavin:1984vu}. For an $(n+1)$-point  correlation function with one degenerate field, these equations are tools to investigate the above-mentioned fundamental objects. First type of objects is the $3$-point structure constants, which are determined in the following way: the $(3+ 1) =4$-point correlation functions satisfy ordinary differential equations on conformal blocks, solving which (with particular monodromy conditions) and using the analytic properties of solutions, we can find the structure constants \cite{Zamolodchikov:1995aa,Teschner:2001rv}. These structure constants completely determine the $3$-point correlation functions.

The second type of objects is  the ($n \geq 4$)-point conformal blocks. These blocks can be straightforwardly written as a series in cross-ratios but remains unknown in the closed form. In various contexts it is interesting to study the classical limit ($c\rightarrow\infty$), assuming that all conformal dimensions grow linearly in $c$. In the limit, the Virasoro blocks have the exponential form (i.e. $F(z) \sim \exp(c/6 f(z))$, where the function $f(z)$ is the classical block), see \cite{Zamolodchikov:1985ie, Besken:2019jyw}. 
One way to find the classical block is to consider the auxiliary $(n+1)$-point conformal block with the additional degenerate operator which satisfies the second-order BPZ equation. By analyzing the monodromy properties of the solutions of the BPZ equation, one can find the accessory parameters \cite{Harlow:2011ny} which yield the classical block. For example, in the case of $n=4$ classical blocks the accessory parameters are constructed using the certain solutions of the Painlevé VI equation \cite{Litvinov:2013sxa}. The $(n>4)$-point case becomes more complicated, but one can use the heavy-light (HL) approximation \cite{Fitzpatrick:2014vua, Alkalaev:2015wia, Alkalaev:2018nik, Alkalaev:2019zhs, Alkalaev:2020kxz, Pavlov:2021lca} to find the perturbative expansion for the classical blocks.

The above analysis can also be carried out in CFT with extended symmetries, in particular, the $\mathcal{W}_N$ symmetry \cite{Fateev:1987vh, Bouwknegt:1994zux}. In this paper, we focus on the $\mathcal{W}_3$  case. Compared with the Virasoro algebra, this case is more complicated due to the fact that the $\mathcal{W}_3$ symmetry is not strong enough and OPE of the primary operators involves descendant operators (associated with $W_3$ current) which can not be expressed in terms of differential operators acting on the primary operators. One can distinguish between non-degenerate, semi-degenerate and fully degenerate representations in $\mathcal{W}_3$ CFT~\cite{Fateev:1987vh, Bowcock:1992gt, Bowcock:1993wq} (the precise definition will be given below). 
The $4$-pt conformal blocks with one fully degenerate and one level-1 semi-degenerate operators are governed by a third-order differential equation (of BPZ-type)~\cite{Fateev:2007ab}. The equations for conformal blocks with higher level semi-degenerate operators are more complicated, e.g. for level-2 semi-degenerate operators one obtains 6-th order ODE \cite{Belavin:2016wlo} (see also \cite{Belavin:2016qaa}). 
  
Regarding the monodromy method, it can also be generalized, but with the following differences. First, the exponentiation of 4-pt W$_3$ blocks in the classical limit has not been proven and has been studied mainly for the blocks with level-1 semi-degenerate operators \cite{Alkalaev:2024knk}. Second, in addition to classifying operators according to the HL approximation, we must further distinguish semi-degenerate operators by their level $N$. This additional structure makes the overall classification more ramified.

 The monodromy method for calculating $\mathcal{W}_3$ blocks was considered in  \cite{deBoer:2014sna}. This consideration focuses on the blocks with non-degenerate primary operators and it implicitly implies that the blocks involving $\mathcal{W}_3$ descendant operators are exponentiated in the classical limit. We note that this assumption is not obvious, since $\mathcal{W}_3$ descendants are not related to the primary operators by differential operators, as in the Virasoro case. Furthermore, the monodromy method in \cite{deBoer:2014sna} was restricted to computing classical blocks with an identity intermediate operator. In this work, we consider a more general case.

In this article, we focus on classical 4-pt blocks with level-1 and level-2 semi-degenerate operators and a non-identity intermediate operator. We derive the BPZ-type equations for the auxiliary 5-pt blocks with  one additional fully degenerate operator. Using the HL approximation, we describe the monodromy properties of the BPZ equation solutions and then derive the monodromy equations for the accessory parameters. Solving these equations explicitly yields the corresponding 4-pt classical blocks.

The paper is organized as follows. In Section \bref{sec:Pre} we recall main facts about the $\mathcal{W}_3$ CFT and introduce the $\mathcal{W}_3$ conformal blocks. In Section \bref{sec:BPZ} we present the BPZ-type equations for the auxiliary 5-pt blocks with level-1 and level-2 semi-degenerate operators and one fully degenerate operator. In Section \bref{sec:HL} we derive and solve the monodromy equations within the HL approximation, which gives us the classical 4-pt blocks we are interested in. Concluding Section \bref{sec:MP} summarizes our results and contains future developments.

\section{Preliminaries}

\label{sec:Pre}

\subsection{The $\mathcal{W}_3$ algebra and its representations}

\paragraph{Algebra.} The $\mathcal{W}_3$  symmetry is generated by the spin-2 and spin-3  holomorphic currents ($T(z)$ and $W(z)$), which can be expanded as follows 
\be
\label{curr}
T(z) = \sum_{m \in \mathbb{Z}}\frac{L_m}{z^{m+2}} \;, \qquad W(z) = \sum_{n \in \mathbb{Z}}\frac{W_n}{z^{n+3}}\;.
\ee
The currents' modes form  the $\mathcal{W}_3$ algebra \cite{Zamolodchikov:1985wn, Fateev:2007ab} 

\be
\ba{c}\label{AW3}
    \left[L_n,L_m\right]=(n-m)L_{n+m}+\dps \frac{c}{12}(n^3-n)
    \delta_{n+m,0}\;,
    \\
    \\
    \left[L_n,W_m\right]=(2n-m)W_{n+m}\;,
    \\
    \\
   \dps \left[W_n,W_m\right]=-\frac{c}{144}(n^2-1)(n^2-4)n
    \delta_{n+m,0}-\frac{40}{22+5c}(n-m)\Lambda_{n+m}-\\
    \\
    \dps\frac{(n-m)}{12}\left( 2m^2 +2n^2 - m n -8  \right)L_{n+m}\;, \qquad n, m \in \mathbb{Z},
\ea
\ee
where\footnote{The algebra \eqref{AW3} differs from the algebra (with generators $\tilde W_n$) in \cite{Fateev:1987vh} by the following redefinition of the generators: $W_n = \tilde{W}_n/(-i \sqrt{5/2})$.}
\be
\label{nl}
\Lambda_{m} = \sum_{p \leq - 1} L_{p} L_{m-p} + \sum_{p \geq 0 } L_{m-p} L_{p} - \frac{3(m+2)(m+3)}{10} L_{m}\;.
\ee
 Also, we will use the Toda parametrization for the central charge
\be
\label{TP}
c = 2 + 24 ( b+b^{-1})^2. 
\ee
\paragraph{HW  representations.} The highest weight (HW) representations are defined in the similarly  to the Virasoro case. There are two Cartan generators ($L_0$ and $W_0$), so the HW representation satisfies 
\be
\ba{c}
\label{HW}
L_0 |h, w\rangle = h |h, w\rangle, \qquad W_0 |h, w\rangle = w |h, w\rangle, \\
L_n |h, w\rangle = 0, \qquad W_m |h, w\rangle = 0, \quad m, n \geq 1. 
\ea
\ee
A$\mathcal{W}_3$  Verma module is spanned by the descendant states 
\be
\label{desc}
\mathcal{L}_{-I} |h, w\rangle \equiv L_{-i_m}...L_{-i_1} W_{-j_n}...W_{-j_1} |h, w\rangle, 
\ee
where $i_m \geq ...\geq i_1 \geq 1$, $j_n \geq ...\geq j_1 \geq 1$ and the level of the descendant state is $|I| \equiv i_1 + i_2 + ...+ j_1 +j_2 +...$ A $\mathcal{W}_3$  primary operator $\cO(z)$, corresponding to the HW representation \eqref{HW}, is characterized by the conformal dimension ($h$) and the spin-3 charge ($w$). The convenient way to parametrize $h$ and $w$ is to use the vector $\vec{\alpha}$ spanning by the fundamental $sl_3$ weights, 
\begin{equation}
\label{W3P}
\vec{\alpha} = \alpha_1 \; \vec{\omega}_1 + \alpha_2 \; \vec{\omega}_2\;,
\end{equation}
where
\be
\vec{\omega}_1 = \sqrt{\frac23} \left(1, 0\right),\quad  \vec{\omega}_2 = \sqrt{\frac23} \left(\frac12, \frac{\sqrt{3}}{2}\right).
\ee
By introducing vectors $\vec{e}_{1,2}, \vec{p}_i$, $i=1,2,3$ and $\vec{\rho}$
\be
\label{W3Pd}
\vec{e}_1 =  2 \vec{\omega}_1 - \vec{\omega}_2,\quad 
\vec{e}_2 =  -\vec{\omega}_1 + 2 \vec{\omega}_2,
\quad
\vec{p}_1 =  \vec{\omega}_1,\quad 
\vec{p}_2 =  \vec{\omega}_2 - \vec{\omega}_1 ,\quad
\vec{p}_3 =  -\vec{\omega}_2, \quad \vec{\rho}  = \vec{\omega}_1 + \vec{\omega}_2, 
\ee
we construct variables $x_i$ in terms of which the conformal dimensions/spin-3 charges can be represented compactly
\begin{equation}
\label{daq}
x_i =  \left((b+b^{-1})\vec{\rho}-\vec{\alpha}\right) \cdot \vec{p}_i: \quad	h  = (b+b^{-1})^2+ x_1 x_2 + x_1 x_3 + x_2 x_3, \quad w =  \frac{-x_1 x_2 x_3}{\sqrt{b^2+\frac{1}{b^2}+\frac{34}{15}}} .
\end{equation}
From \eqref{W3Pd} it follows that  $x_1 + x_2 + x_3 =0$. Another observation is that swapping $\vec{\omega}_1 \leftrightarrow \vec{\omega}_2$ corresponds to $h\rightarrow h$, $w\rightarrow -w$. Finally, if $\vec{\alpha}$ does not explicitly depend on $b$ then $h$ and $w$ are invariant under the change $b\rightarrow1/b$. 

\paragraph{Semi- and fully degenerate representations.}
For the specific  $\vec{\alpha}$ the corresponding representation is reducible.  There are two distinct situations: semi-degenerate representations and fully degenerate representations, which we describe here. 

Focusing on generic values of $b$ (thereby excluding the $\mathcal{W}_3$ minimal models), these semi-degenerate representations exhibit a single null vector at level $M\times N$ (see \cite{Fateev:1987vh, Bowcock:1992gt}). For such representations, the vector $\vec{\alpha}$ has the form (with an analogous expression holding for representations related by the exchange  $\vec{\omega}_1 \leftrightarrow \vec{\omega}_2$)
\be
\label{semi-deg}
\vec{\alpha} = \left(b (1 - M)+ \dps \frac{1-N}{b}\right) \vec{\omega}_1 + s \vec{\omega}_2, \qquad M, N \in \mathbb{Z}_{+}, \qquad s \in \mathbb{R}.
\ee
In what follows, we consider level-1 ($M=1, N=1$) and level-2 ($M=1, N=2$) semi-degenerate representations. It follows from \eqref{semi-deg}, that in the classical limit ($b\rightarrow0$) the first term in brackets disappears for any $M$. Therefore, for another semi-degenerate level-2 representation ($N=2, M=1$) in the classical limit the vector $\vec{\alpha}$ \eqref{semi-deg} coincides with the level-1 representations. 

The level-1 semi-degenerate representation corresponds to  $\vec{\alpha} = s \vec{\omega}_2$ and has one null-vector 
\be
\label{Level.1.identity}
\Psi_{s\vec{\omega}_2}= \left( W_{-1} - \frac{3 w}{2h} L_{-1} \right)  \cO_{s\vec{\omega}_2},
\ee
where the conformal dimension/spin-3 charge $h$ and $w$ are
\be
\label{hw_l1}
h = \frac{s \left(3 b^2-b s+3\right)}{3 b}, \qquad w = - \frac{s \left(3 b^2-2 b s+3\right) \left(3 b^2-b s+3\right)}{27 b^2 \sqrt{b^2+\frac{1}{b^2}+\frac{34}{15}}}. 
\ee

The level-2 semi-degenerate representation $\vec{\alpha} = s \vec{\omega}_2 - \vec{\omega}_1/b$ has following null-vector 
\be
\label{N2}
\Psi_{\vec{\alpha}} =  \left(
c_{1}  L_{-2} +
c_{2}  W_{-2}   +
c_{3}  L_{-1}^2 + 
c_{4}  L_{-1} W_{-1} + W_{-1}^2 \right) \cO_{\vec{\alpha}},
\ee
where
\be
\label{Level2s}
\ba{c}
\dps c_1 = \frac{5 \left(b^2-b s+2\right) \left(2 b^2-b s+2\right)}{b^2 \left(3 b^2+5\right) \left(5 b^2+3\right)}, \qquad c_2 = \frac{\left(3 + b^2\right) \left(4-2 b s+3 b^2\right)}{2 b^2 \sqrt{9 b^4+\frac{102 b^2}{5}+9}}, \\
\\ \dps c_3 = \frac{5 \left(3 b^2-2 b s+1\right) \left(3 b^2-2 b s+7\right)}{12 \left(3 b^2+5\right) \left(5 b^2+3\right)}, \qquad c_4 = \frac{4 b^2-2 b s+4}{\sqrt{9 b^4+\frac{102 b^2}{5}+9}}.
\ea
\ee
The conformal dimension/spin-3 charge for the level-2 semi-degenerate representation are
\be
\label{hw_l21b}
h = \frac{3 b^3 s-b^2 \left(s^2+3\right)+4 b s-4}{3 b^2}, \qquad w = -\frac{\left(3 b^2-2 b s+4\right) \left(3 b^2-b s+5\right) (b s+1)}{9 b^2 \sqrt{9 b^4+\frac{102 b^2}{5}+9}}.
\ee
For the construction of the BPZ-type equations, we utilize the fully degenerate representation with a weight vector, $\vec{\alpha} = -b \vec{\omega}_1$, which possesses three singular vectors \cite{Fateev:1987vh, Fateev:2007ab}
\be 
\label{deg}
    \Psi_1  =  \left(W_{-1}-\frac{3q_{\psi}}{2 h_{\psi}}L_{-1}\right) \cO_{-b \vec{\omega}_1},
    \ee
    \be
    \Psi_2  = \left(W_{-2}-\frac{6w_{\psi}}{h_{\psi}(5h_{\psi}+1)}\left( L_{-1}^2 - (h_{\psi}+1)L_{-2}\right) \right)\cO_{-b \vec{\omega}_1},
\ee
\be\label{3null}
 \Psi_3 = \left(W_{-3}+ \frac{w_{\psi}}{h_{\psi}(5h_{\psi}+1)} \left( -
  \frac{16}{h_{\psi}+1}
    L_{-1}^3+ 12 L_{-1}L_{-2}+
    \frac{3(h_{\psi}-3)}{2} L_{-3}\right)\right) 
   \cO_{-b \vec{\omega}_1},
\ee
where
\be
\label{resc}
\dps h_{\psi} = -1 - \frac{4 b^2}{3}, \qquad w^2_{\psi} =  \frac{5 h^2_{\psi}}{27} \frac{5 b + 3 b^{-1}}{3 b + 5 b^{-1}}.
\ee

\paragraph{Classical limit.} 
The classical limit $c \rightarrow \infty$  can be realized as $b\rightarrow0$. We are interested in the operators characterized by the following behavior in the classical limit
\be
\label{classical_parameters}
\text{at} ~~ b \rightarrow 0: \quad \epsilon = \frac{h b^2}{4}, \qquad q = \frac{w b^2}{4} \quad \text{are finite}, 
\ee
where $\epsilon, q$ are called classical dimensions/spin-3 charges.  For a non-degenerate operator, the classical dimensions and spin-3 charges are expressed in terms of auxiliary parameters $\tilde{x}_i$ as follows
\be
\label{top}
4 \epsilon = 1 + \tilde x_1 \tilde x_2 + \tilde x_3 \tilde x_2 + \tilde x_1 \tilde x_3, \quad  4 q =  - \; \tilde x_1  \tilde x_2  \tilde x_3, \qquad \text{where} ~~~ \tilde x_i = x_i  b, ~~ b\rightarrow 0. 
\ee 
Introducing an auxiliary parameter $a$, according to 
\be
\text{at} \quad b \rightarrow 0: \qquad s = \frac{a}{b},  \quad a ~ \text{is finite}, 
\ee 
the classical dimensions and spin-3 charges for level-1 operators can be  written as follows 
\be
\label{classical_d_c1}
\epsilon^{(1)} = \frac{a - a^2/3}{4}, \qquad q^{(1)} = \frac{ (a-3) a (2 a -3)}{108},    
\ee
and for the level-2 operators~\eqref{hw_l21b} one has
\be
\label{classical_d_c2}
\epsilon^{(2)}=-\frac{1}{12} (a-2)^2, \qquad q^{(2)} = \frac{1}{54} (a-5) (2 - a) (a+1). 
\ee
The parameters $\tilde x_i$ in \eqref{top} for the level-1 and level-2 semi-degenerate operators $(N=1,2)$ are
\be
\label{HRN}
\tilde x_1 = -\frac{1}{3} (1-a+2 N), \qquad \tilde x_2 = \frac{1}{3} (a-1+N), \qquad \tilde x_3 = \frac{1}{3} (2-2 a+N). 
\ee
Explicit forms of the level-1 \eqref{Level.1.identity}  and level-2 \eqref{N2} null-vectors in the classical limit are  
\be
\label{L1Classical}
\left( W_{-1} - \frac{(2 a-3)}{6}  L_{-1} \right)  \cO_{\frac{a \vec{\omega}_2}{b}} =0, 
\ee
and
\be
\label{L2ClassicalS}
\left(  \frac{(a-2)^2}{3 b^2}  L_{-2} +
\frac{2-a}{b^2} W_{-2}  + \frac{\left(4 a^2-16 a+7\right)}{36}L_{-1}^2 - \frac{2(a-2) }{3}  L_{-1} W_{-1} + W_{-1}^2 \right) \cO_{\frac{a \vec{\omega}_2 -  \vec{\omega}_1}{b}} =0.
\ee

\subsection{Conformal blocks and the Ward identities in $\mathcal{W}_3$ CFT}

To construct the blocks, one can define a (block-diagonal) Shapovalov matrix 
\be
\label{sh}
H^{IJ} = \langle \tilde h_p, \tilde w_p  | \mathcal{L}_{J} \mathcal{L}_{-I}| \tilde h_p, \tilde w_p \rangle,
\ee
where the matrix elements of $H^{IJ}$ are scalar products of descendant states \eqref{desc}. Matrix elements of the primary operator $\cO(z)$ have the following form
\be
\label{tp}
\Gamma'_{I}(\vec{\tilde \alpha}_p, \vec{\alpha},  \vec{\alpha}_p) = \frac{\langle \tilde h_p, \tilde w_p  | \mathcal{L}_{I} ~ \cO(1)|h_p,  w_p \rangle}{ \langle \tilde h_p, \tilde w_p  |\cO(1)| \tilde h_p, \tilde w_p \rangle}, \quad \Gamma_{I} ( \vec{\alpha}_p, \vec{\alpha}, \vec{\tilde \alpha}_p) = \frac{\langle  h_p, w_p  | \cO(1)  \mathcal{L}_{-I} |\tilde h_p, \tilde w_p  \rangle}{ \langle \tilde h_p, \tilde w_p  |\cO(1)| \tilde h_p, \tilde w_p \rangle}. 
\ee
A $4$-pt $\mathcal{W}_3$ conformal block is a function of coordinates $z_i$ and conformal dimensions/spin-3 charges $h_i, w_i$, $i = \overline{1,4}$ which we denoted as $F_4(z_i| h, w)$ 
\be
\ba{c}
F_4(z_i| h, w) = \left(z_4-z_1\right){}^{-2 h_4} \left(z_4-z_2\right){}^{h_1-h_2+h_3-h_4} \left(z_4-z_3\right){}^{h_1+h_2-h_3-h_4} \left(z_3-z_2\right){}^{h_4-h_1-h_2-h_3} \times \\
\\\dps F\left(\eta| h, w \right), \quad \eta = \frac{(z_2 - z_3)(z_1 - z_4)}{(z_1 - z_3)(z_2 - z_4)}. 
\ea
\ee
where $ \dps F(\eta| h, w)$ is defined via  \eqref{sh} and \eqref{tp}
\be
\label{B}
F(\eta| h, w) = \eta^{\tilde h_p - h_1 - h_2}\sum^{\infty}_{k=0} \eta^k \sum_{\substack{
K, K'  
\\ 
|K| = |K'| = k 
}
} \Gamma_{K}(\vec{\alpha}_1, \vec{\alpha}_2, \vec{\tilde \alpha}_p) \left(H\right)^{-1}_{K K'} \Gamma'_{K'}(\vec{\tilde \alpha}_p, \vec{\alpha}_3, \vec{\alpha}_4). 
\ee
The multi-point $\mathcal{W}_3$ conformal blocks can be defined in the same fashion. The $n$-point $\mathcal{W}_3$ conformal blocks satisfy 8 Ward identities, 3 of which are associated with the current $T(z)$
\be
\label{stY}
\ba{c}
\dps \sum^{n}_{i=1} \partial_{i} F_n(z_i|h, w) =0, \\
\dps \sum^{n}_{i=1} (z_i \partial_{i} + h_i)  F_n(z_i|h, w)  =0, \\
\dps \sum^{n}_{i=1} (z^2_i \partial_{i} + 2 z_i h_i) F_n(z_i|h, w) =0, 
\ea
\ee
and there are 5 Ward identities, associated with the current $W(z)$
 \be
 \label{extY}
 \ba{c}
 \dps \sum^{n}_{i=1} \mathcal{W}_{_{-2}}^{(i)} F_n(z_i|h, w) =0, \\
  \dps \sum^{n}_{i=1} \left(z_i \mathcal{W}_{_{-2}}^{(i)} +  \mathcal{W}^{(i)}_{_{-1}} \right) F_n(z_i|h, w)=0, \\
 \dps \sum^{n}_{i=1} \left(z^2_i \mathcal{W}_{_{-2}}^{(i)} + 2 z_i \mathcal{W}^{(i)}_{_{-1}} + w_i\right) F_n(z_i|h, w)=0, \\
  \dps \sum^{n}_{i=1} \left(z^3_i \mathcal{W}_{_{-2}}^{(i)} + 3 z^2_i \mathcal{W}^{(i)}_{_{-1}} + 3 z_i w_i\right) F_n(z_i|h, w)=0, \\
   \dps \sum^{n}_{i=1} \left(z^4_i \mathcal{W}_{_{-2}}^{(i)} + 4 z^3_i \mathcal{W}^{(i)}_{_{-1}} + 6 z^2_i w_i\right) F_n(z_i|h, w)=0,
   \ea
   \ee
where the notation $\mathcal{W}_{_{-1,-2}}^{(i)} F_n(z_i|h, w)$ stands for a conformal block where the operator $\cO(z_i)$ is replaced by $\mathcal{W}_{_{-1,-2}}^{(i)} \cO(z_i)$.

In the classical limit, the 4-pt $\mathcal{W}_3$ block \eqref{B} takes the exponential form \cite{Fateev:2007ab, Alkalaev:2024knk}
\be
\label{classicalf}
F (\eta| h, w)\Big|_{b \rightarrow 0} \rightarrow \exp\left( \frac{4 f(\eta| \epsilon, q)}{b^2}\right),
\ee
where the classical dimensions/spin-3 charges of all (external and internal operators) are given by \eqref{classical_parameters}.

\section{BPZ-type equations}

\label{sec:BPZ}

We consider the 4-pt classical blocks with the semi-degenerate operators. Within the monodromy method, these blocks can be extracted from the monodromy properties of auxiliary $5$-pt blocks (denoted by  $\Psi (y, z_i) $) with one additional fully degenerate operator $\cO_{-b \vec{\omega}_1}(y)$. 

The block $\Psi (y, z_i)$ satisfies the BPZ-type  equation, which follows from the null-vector \eqref{3null}, where all generators $(L, W)$ are rewritten as differential operators. According to the Ward identities, action of $W_{-3}$ on $\Psi(y, z_i)$ takes the form 
\be
\label{w3}
W_{-3} \Psi (y, z_i) = \sum^{4}_{i=1} \left( \frac{w_i}{(y-z_i)^3} + \frac{\mathcal{W}^{(i)}_{-1}}{(y-z_i)^2} + \frac{\mathcal{W}^{(i)}_{-2}}{(y-z_i)} \right) \Psi (y, z_i). 
\ee
The 8 functions  $\mathcal{W}^{(i)}_{-1,-2} \Psi (y, z_i), ~ i = \overline{1,4}$ are related with  $\Psi (y, z_i)$ by 5 Ward identities \eqref{extY}. In the case of 4 non-degenerate operators the BPZ-type equation involves  3 independent functions.

The blocks with semi-degenerate operators are subject of additional (to the Ward identities) constraints.  We consider two distinct cases:  1) blocks with 3 level-1 operators and 2) blocks with 2 level-1 and 1 level-2 operators. The blocks with multiple level-2 operators represent a more complicated situation which will be analyzed elsewhere. For both of the considered cases it follows from \eqref{classicalf} that the corresponding auxiliary $5$-pt blocks have the exponential behavior \eqref{classicalf} with a certain prefactor (the fully degenerate operator does not affect the leading behavior due to $h_{\psi} \ll h_i, ~ w_{\psi} \ll w_i$ at $b\rightarrow0$)
\be
\label{EF}
 \Psi (y, z_i) \Big|_{b \to 0} \rightarrow \psi(y,z_i) \exp\left[\frac{4}{b^2}f (z_i)\right]\;.
\ee
Moreover, all conformal blocks $\mathcal{W}^{(i)}_{-1,-2} \Psi (y, z_i), ~ i = \overline{1,4}$ have the same behavior in the classical limit 
\be\label{CL}
\mathcal{W}^{(i)}_{-1,-2} \Psi (y, z_i) \Big|_{b \rightarrow 0} \rightarrow \frac{k^{(i)}_{1,2}(y, z_i)}{b^2}\exp\left[\frac{4}{b^2}f (z_i)\right], \qquad i = \overline{1,4}, 
\ee
where $k^{(i)}_{1,2}$ are some $b$-independent functions. Indeed, if we consider the 3 level-1 semi-degenerate operators at points $z_{p}$ where  $p = i_1, i_2, i_3$, then
\be
\mathcal{W}^{(p)}_{-1} \Psi (y, z_i) = \frac{3 w_p}{2 h_p} \partial_{z_p} \Psi (y, z_i),
\ee
which is consistent with \eqref{CL}. Remaining conformal blocks $\mathcal{W}^{(i)}_{-2} \Psi (y, z_i)$ and $\mathcal{W}^{(i_4)}_{-1} \Psi (y, z_i)$ are linear combination of spin-3 charges $w_i$ and  the blocks $\mathcal{W}^{(p)}_{-1} \Psi (y, z_i)$ due to the Ward identities \eqref{extY}. Hence, in the classical limit they have the form \eqref{CL}. 

The same holds for the auxiliary blocks (denoted by $\tilde \Psi (y, z_i)$) with one level-2 semi-degenerate operator $\cO(z_i)$ and two level-1 semi-degenerate operators $\cO(z_p)$, $p= i_1, i_2$ in the classical limit, which satisfy \eqref{L2ClassicalS} 
\be
\label{R2}
\left(  \frac{(a_i-2)^2}{3 b^2}  \mathcal{L}^{(i)}_{-2} +
\frac{2-a_i}{b^2} \mathcal{W}^{(i)}_{-2}  + \frac{\left(4 a_i^2-16 a_i+7\right)}{36} \partial^2_{z_i}  - \frac{2(a_i-2)}{3}  \partial_{z_i} \mathcal{W}^{(i)}_{-1} + \left( \mathcal{W}_{-1}^{(i)} \right)^2 \right) \tilde \Psi (y, z_i)=0.  
\ee
One can see that in the classical limit the behavior of the functions \eqref{CL} is consistent (taking into account Ward identities) with this condition \eqref{R2}. 

\subsection{Blocks with one non-degenerate and 3 level-1 semi-degenerate operators}
\label{Sec:BPZ1}

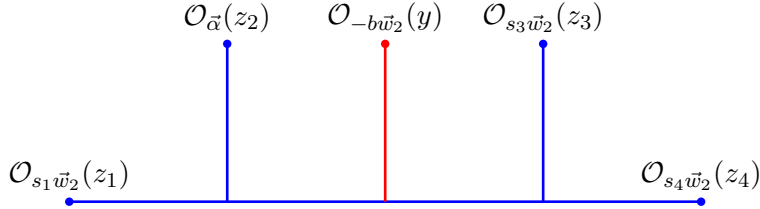
\begin{figure}[H]
\centering
\begin{tikzpicture}[scale=0.70]

\draw [blue,decorate, line width=1pt] (35,0) -- (35,3);

\draw [blue,line width=1pt] (32,0) -- (38,0);
\draw [blue, line width=1pt] (41,0) -- (41,3);
\draw [blue, line width=1pt] (38,0) -- (44,0);
\draw [red, line width=1pt] (38,0) -- (38,3);


\draw (32.0, 0.5) node {$\cO_{s_1 \vec{\omega}_2} (z_1)$};
\draw (35.0,3.5) node {$\cO_{\vec{\alpha}} (z_2)$};
\draw (41.0,3.5) node {$\cO_{s_3 \vec{\omega}_2} (z_3)$};
\draw (44, 0.5) node {$\cO_{s_4 \vec{\omega}_2} (z_4)$};

\draw (38.0,3.5) node {$\cO_{-b \vec{\omega}_2}(y)$};


\fill[blue] (35,3) circle (0.8mm);

\fill [blue]      (32,0)  circle (0.8mm);;


\fill[red] (38,3) circle (0.8mm);

\fill   [blue]    (41,3) circle (0.8mm);;

\fill    [blue]   (44,0) circle (0.8mm);;

\end{tikzpicture}
\label{N1S4}
\caption{The auxiliary $5$-pt block  $\Psi (y, z_i)$ for the $4$-pt block with one non-degenerate operator and 3 level-1 degenerate operators. The fully degenerate operator is depicted in red. }
\end{figure}

We consider the 4-pt classical block with one non-degenerate operator at $z_2$ and three level-1 semi-degenerate operators, located at $z_1$, $z_3$ and $z_4$. An auxiliary $5$-pt block  $\Psi (y, z_i)$ for such a case depicted in Fig. \bref{N1S4}. 

By construction, the auxiliary block satisfies three following conditions
\be
\label{B3N1}
\mathcal{W}^{(p)}_{-1} \Psi (y, z_i) = \frac{3 w_p}{2 h_p} \partial_{z_p} \Psi (y, z_i), \qquad p = 1, 3, 4,   
\ee
and another one, associated with the condition \eqref{3null}
\be
\label{new3}
 \left(W_{-3}+ \frac{w_{\psi}}{h_{\psi}(5h_{\psi}+1)} \left( -
  \frac{16}{h_{\psi}+1}
    \partial_y^3+ 12 \partial_y L^{(y)}_{-2}+
    \frac{3(h_{\psi}-3)}{2} L^{(y)}_{-3}\right)\right) \Psi (y, z_i) = 0. 
    \ee

To rewrite \eqref{new3} as the BPZ-type equation, we do the following steps:
\begin{enumerate}
    \item  Express operators $L^{(y)}_{-3}$ and $L^{(y)}_{-2}$ in \eqref{new3} as differential operators with respect to $y$ and $z_i$.
    \item Solve the Ward identities \eqref{extY}, expressing $\mathcal{W}^{(i)}_{-2} \Psi (y, z_i), ~ i = \overline{1,4}$ and $\mathcal{W}^{(2)}_{-1} \Psi (y, z_i)$ through $\mathcal{W}^{(p)}_{-1} \Psi (y, z_i)$, $p=1,3,4$ and $\Psi (y, z_i)$. Then, use identities \eqref{B3N1} to rewrite $W_{-3} \Psi (y, z_i)$ \eqref{w3} as a differential operator, acting on $\Psi (y, z_i)$. 
    \item Choose $(y, z_1, z_2, z_3, z_4) \rightarrow (x,1,z,0,\infty)$ for convenience. 
    \item Take the classical limit, applying  \eqref{classical_parameters}  and \eqref{EF} to the conformal block $\Psi (y, z_i)$. 
    \end{enumerate} 
The resulting BPZ-type equation takes the form 
\be\label{BPZ}
\left[\frac{d^3}{dx^3} + 4 T(x,z) \frac{d}{dx} + \frac{2 d T(x,z)}{dx} - 4W(x, z) \right]\psi(x, z) =0\;,
\ee
where  $T(x, z)$ reads
\be
\label{TG}
T(x,z) = \frac{\epsilon_1}{(x-1)^2} + \frac{\epsilon_2}{(x-z)^2} + \frac{\epsilon_3}{x^2} + \frac{(z-1) z c(z)}{(x-1) x (x-z)} - \frac{(\epsilon_1+\epsilon_2 +\epsilon_3 - \epsilon_4)}{(x-1) x}, \quad c(z) = \frac{d f (z)}{dz}. 
\ee

The expression for $W(x,z)$ is too bulky for the general choice of classical dimensions and spin-3 charges, so we choose $\epsilon_4 = \epsilon_3 \equiv \epsilon_H$, $q_4 = - q_3 \equiv q_H$\footnote{A similar symmetric configuration involving two identical heavy dimensions was analyzed for the $4$-pt conformal block; this case was shown to be dual to a BTZ background \cite{deBoer:2014sna, Alkalaev:2015wia}.} 
\be
\label{N1W}
\ba{c}
\dps W(x,z) = \frac{q_H}{x^3}- \frac{3 \epsilon_2 q_1 \left(z^2-2 z+1\right)}{2 \epsilon_1 (x-1)^2 x (x-z)^2}  -\frac{3 q_H z (z-1) c(z)}{2 \epsilon_H (x-1) x^2 (x-z)}-\frac{3 q_1 z (z-1)^2c(z)}{2 \epsilon_2 (x-1)^2 x (x-z)^2} 
\\
\\
\dps + \frac{q_2 z(z-1)}{x(x-1)(x-z)^3}-\frac{q_1 (z-1) (x z+3 x-3 z-1)}{2 (x-1)^3 x (x-z)^2} + \frac{3 q_H \left(\epsilon_1 \left(z^2-x\right)+\epsilon_2 \left(-2 x z+x+z^2\right)\right)}{2 \epsilon_H (x-1) x^2 (x-z)^2}. 
\ea
\ee

\subsection{Blocks with one non-degenerate, 2 level-1 and one level-2 operators}

\label{Sec:BPZ22}

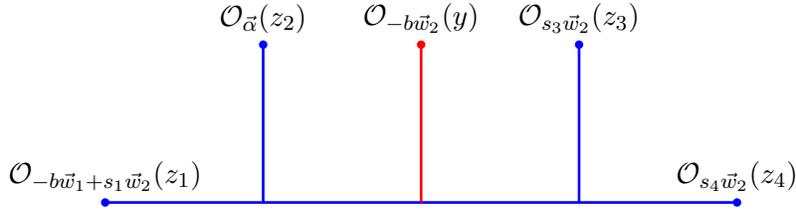
\begin{figure}[H]
\centering
\begin{tikzpicture}[scale=0.70]

\draw [blue,decorate, line width=1pt] (35,0) -- (35,3);

\draw [blue,line width=1pt] (32,0) -- (38,0);
\draw [blue, line width=1pt] (41,0) -- (41,3);
\draw [blue, line width=1pt] (38,0) -- (44,0);
\draw [red, line width=1pt] (38,0) -- (38,3);


\draw (32.0, 0.5) node {$\cO_{- b \vec{\omega}_1+ s_1 \vec{\omega}_2} (z_1)$};
\draw (35.0,3.5) node {$\cO_{\vec{\alpha}}(z_2)$};
\draw (41.0,3.5) node {$\cO_{s_3 \vec{\omega}_2}(z_3)$};
\draw (44, 0.5) node {$\cO_{s_4 \vec{\omega}_2}(z_4)$};

\draw (38.0,3.5) node {$\cO_{-b \vec{\omega}_2}(y)$};


\fill[blue] (35,3) circle (0.8mm);

\fill [blue]      (32,0)  circle (0.8mm);;


\fill[red] (38,3) circle (0.8mm);

\fill   [blue]    (41,3) circle (0.8mm);

\fill    [blue]   (44,0) circle (0.8mm);

\end{tikzpicture}
\label{N2S4}
\caption{The auxiliary $5$-pt block  $\tilde \Psi (y, z_i)$ for the $4$-pt block with one non-degenerate operator, two level-1 semi-degenerate operators and one level-2 semi-degenerate operator. The fully degenerate operator is in red.}
\end{figure}

Consider a 4-pt classical block with 2 level-1 semi-degenerate operators, located at $z_3$ and $z_4$, one level-2 semi-degenerate operator at $z_1$ and one non-degenerate operator at $z_2$. A corresponding auxiliary $5$-pt block is presented in Fig. \bref{N2S4}. 

The block $ \tilde \Psi (y, z_i)$  satisfies the following conditions: 
\be
\label{B2N1a}
\mathcal{W}^{(p)}_{-1} \tilde \Psi (y, z_i) = \frac{3 w_p}{2h_p} \partial_{z_p} \tilde \Psi (y, z_i), \qquad p = 3, 4, 
\ee
the level-2 condition 
\be
\label{B2N2}
 \left( \frac{(a_1-2)^2}{3 b^2}  \mathcal{L}^{(1)}_{-2} +
\frac{2-a_1}{b^2} \mathcal{W}^{(1)}_{-2}  + \frac{\left(4 a_1^2-16 a_1+7\right)}{36} \partial^2_{z_1} - \frac{2(a_1-2) }{3}  \partial_{z_1} \mathcal{W}^{(1)}_{-1} + \left(\mathcal{W}^{(1)}_{-1}\right)^2  \right)\tilde \Psi (y, z_i)=0,
\ee
and
\be
\label{new32}
 \left(W_{-3}+ \frac{w_{\psi}}{h_{\psi}(5h_{\psi}+1)} \left( -
  \frac{16}{h_{\psi}+1}
    \partial_y^3+ 12 \partial_y L^{(y)}_{-2}+
    \frac{3(h_{\psi}-3)}{2} L^{(y)}_{-3}\right)\right) \tilde \Psi (y, z_i) = 0. 
    \ee
The derivation of the BPZ-type equation differs from the derivation in the previous subsection. In a present case, it is more convenient to substitute \eqref{B2N1a} and \eqref{new32} to \eqref{B2N2} (see also \cite{Belavin:2016qaa,Belavin:2016wlo}). Namely, we perform the following steps:
\begin{enumerate}
   \item Express the first term in \eqref{B2N2} via a differential operator, acting on $\tilde \Psi (y, z_i)$ (notice that the third term is already written in such a form).
   \item Use the Ward identities \eqref{extY}, conditions \eqref{B2N1a} and the condition \eqref{new32} to express 
   $\mathcal{W}^{(1)}_{-2} \tilde \Psi (y, z_i), \mathcal{W}^{(1)}_{-1}  \tilde \Psi (y, z_i)$ in terms of differential operators acting on  $\tilde \Psi (y, z_i)$. 
    \item Take the classical limit \eqref{classical_parameters} and \eqref{EF}. Notice that $\mathcal{W}^{(1)}_{-1}  \tilde \Psi (y, z_i) = \tilde k^{(1)}_1 (y, z_i) \tilde \Psi (y, z_i)$ in the classical limit. Hence, the fifth term in \eqref{B2N2} is $\left(\mathcal{W}^{(1)}_{-1}\right)^2 \tilde \Psi (y, z_i) = \left( \tilde  k^{(1)}_1 (y, z_i) \right)^2 \tilde \Psi (y, z_i)$. 
    
    \item Substitute the resulting expressions for $\left(\mathcal{W}^{(1)}_{-1}\right)^2 \tilde \Psi (y, z_i)$,  $\mathcal{W}^{(1)}_{-2} \tilde \Psi (y, z_i)$ and $\mathcal{W}^{(1)}_{-1}  \tilde \Psi (y, z_i)$ into \eqref{B2N2} and choose $(y, z_1, z_2, z_3, z_4) \rightarrow (x,1,z,0,\infty)$. 
\end{enumerate}

The resulting sixth-order equation is too complex to write out in full, but it can be studied using the HL approximation. This equation depends on the accessory parameter (which we denote $\tilde c(z)$) related to the corresponding $4$-pt classical block. We perform a perturbative consideration of this BPZ-type equation in the HL approximation in the Section \bref{sec:HL2}. 

\section{The monodromy method and the HL perturbation theory} 
\label{sec:HL}

Solutions $\psi(x,z)$ of the BPZ-type equations from  sections \bref{Sec:BPZ1} and \bref{Sec:BPZ22} must have a specific monodromy, following from the properties of the fully degenerate representation \eqref{3null}. Moving  around a cycle $\Gamma$ that encloses $1$ and $z$, the solutions transform as $\psi_a (\Gamma \circ x,z) = \tilde{M}_{ab} ~ \psi_b (x,z)$, where the monodromy matrix in the classical limit has a form \cite{Pavlov:2022irx}
\be
\label{MTC}
\tilde{M}_{ab} =  \begin{pmatrix}
  1&0& 0\\
  \\
0&  e^{2\pi i \sqrt{1- 4\epsilon_{p}}}&  0\\
 \\
 0& 0& e^{-2\pi i \sqrt{1- 4\epsilon_{p}}}
\end{pmatrix} \,, 
\ee
where $\epsilon_p$ is a classical dimension of the intermediate operator. Such a condition is sufficient to determine the accessory parameters $c(z)$ and $\tilde c(z)$ in the considered equations. 

The solutions of the equations can not be written explicitly, so we solve them using  the heavy-light (HL) approximation, where a subset of the operators ("heavy") have significantly larger classical dimensions and spin-3 charges than the others ("light"). We focus on the case of two equivalent heavy operators $\epsilon_3 = \epsilon_4 \equiv \epsilon_H, ~ q_3 = - q_4 \equiv q_H$. The HL approximation can be formulated in terms of parameters $\tilde x_i$ through \eqref{top} 
\be
\tilde{x}_i \rightarrow 0^{-} \quad  \text{and} \quad \tilde{x}_{i-1} \tilde{x}_{i+1} \rightarrow -1: \quad \epsilon \ll 1, \quad q \ll 1. 
\ee 
For the level-1 semi-degenerate operators \eqref{classical_d_c1} the HL approximation corresponds to
\be
\label{HLd1}
 \epsilon \ll \epsilon_{H}, \qquad q\ll q_{H},   
\ee 
 with the additional constraint  $\epsilon = - 3 q$. For the level-2 semi-degenerate operators \eqref{classical_d_c2} the HL approximation is
\be
\label{HLd2}
 \epsilon \ll \epsilon_{H}, \qquad q \ll q_{H},
\ee
with the constraint $\epsilon \ll q$. 
\subsection{The monodromy equation for the 4-pt block with one non-degenerate and 3 level-1 operators}
\label{sec:HL1}
Consider the BPZ-type equation \eqref{BPZ} and assume that two level-1 semi-degenerate operators located at points $z_3=0$ and $z_4=\infty$ are heavy, while remaining two external and one intermediate operators are light 
\be
\label{HL}
\epsilon_{1,2}, \epsilon_p \ll \epsilon_H, \qquad q_{1,2} \ll q_H.
\ee
The HL expansion reads
\be
\ba{c}
\label{EX}
  \psi(x,z) = \psi^{(0)}(x) +  \psi^{(1)}(x,z) +...,  \quad T(x,z) = T^{(0)}(x) +  T^{(1)}(x,z) +...,  \\
  \\
  W(x,z) = W^{(0)}(x) +  W^{(1)}(x,z) +..., \quad c(z) = c^{(1)} (z) + .... 
\ea
\ee
Notice that the expansion for the accessory parameter starts from the first order, because the zeroth order classical block is a constant and $c^{(0)} = 0$. Here and below we denote $c^{(1)}(z)$ by $c(z)$, since we consider only the first order in the HL approximation. 

Applying \eqref{EX} to \eqref{BPZ}, in the zeroth order we get
\be
\label{0o}
D^{(0)} \psi^{(0)}(x) =0, \quad  D^{(0)} = \left[\frac{d^3}{dx^3} + 4 T^{(0)}(x) \frac{d}{dx} + \frac{2 d T^{(0)}}{dx} - 4 W^{(0)}(x) \right], 
\ee
where
\be
  T^{(0)}(x) = {\epsilon_4\over x^2}, \quad  W^{(0)}(x) =  \frac{q_H}{x^3}.  
\ee
In the first order the equation \eqref{BPZ} reads
\be
\label{BPZ1}
D^{(0)}  \psi^{(1)}(x,z) = - D^{(1)}  \psi^{(0)}(x)\;, \qquad D^{(1)} =  \left[4 T^{(1)} \frac{d}{dx} + \frac{2 d T^{(1)}}{dx} - 4W^{(1)}\right]\;,
\ee
where 
\be
\label{fot}
T^{(1)}(x,z)  =  \frac{(z-1) c(z)+\epsilon_1+\epsilon_2}{x}+\frac{z c(z)+ \epsilon_1 + \epsilon_2}{1-x}+\frac{c(z)}{x-z}+\frac{\epsilon_2}{(x-z)^2}+\frac{\epsilon_1}{(x-1)^2} \;,
\ee
\be
\ba{c}
\label{fow}
\dps W^{(1)}(x,z ) =  \frac{\epsilon_2 z(z-1)}{3x(x-1)(x-z)^3}  -\frac{3 q_H z (z-1) c(z)}{2 \epsilon_H (x-1) x^2 (x-z)}-\frac{\epsilon_1 z (z-1)^2c(z)}{2 \epsilon_2 (x-1)^2 x (x-z)^2} 
\\
\\
\dps -\frac{ \epsilon_2 \left(z^2-2 z+1\right)}{2 (x-1)^2 x (x-z)^2}-\frac{\epsilon_1 (z-1) (x z+3 x-3 z-1)}{6 (x-1)^3 x (x-z)^2} + \frac{3 q_H \left(\epsilon_1 \left(z^2-x\right)+\epsilon_2 \left(-2 x z+x+z^2\right)\right)}{2 \epsilon_H (x-1) x^2 (x-z)^2}.
\ea
\ee
Solutions of the  equation \eqref{0o} are 
\be 
\psi^{(0)}_{i} = x^{1+ p_i}~,\quad i =1,2,3~,
\ee
where $p_i$ are roots of the cubic equation $p^3 -(1-4 \epsilon_H)p-4 q_H =0$. They are expressed via  parameter $a_H$ using \eqref{top} and \eqref{classical_d_c1} as 
\be
\label{rc}
p_1 = a_H/3, \qquad p_2 = (a_H-3)/3, \qquad p_3 = (3-2a_H)/3. 
\ee
The first-order solutions of \eqref{BPZ1} can be expressed in terms of zeroth-order solutions (here $p_{ij} = p_i - p_j$)
\be
\label{s1}
\psi^{(1)}_{i} (x,z) = \psi_j^{(0)}(x) \int T_{ij}(y, z) \; dy\;,
\ee
where
\be
\label{mf1}
T_{ij} (x,z) = 2 \begin{pmatrix}
  \dps \frac{ 2 x \left( x W^{(1)} -  p_1  T^{(1)}\right)}{ p_{12} p_{13}} \;\;& \dps \frac{x^{1 - p_{12}} \left(p_3 T^{(1)} + 2 W^{(1)} x\right)}{p_{12} p_{13}} \;\;& \dps \frac{ x^{1 - p_{13}} \left(p_2 T^{(1)}+2 W^{(1)} x\right)}{p_{12} p_{23}}\\
  \\
\dps -\frac{ x^{1 + p_{12}} \left(p_3 T^{(1)}+2 W^{(1)} x\right)}{p_{12} p_{23}}\;\;&  \dps - \frac{ 2 x\left(  x W^{(1)} -  p_2 T^{(1)}\right)}{ p_{12} p_{13}} \;\;& \dps -\frac{ x^{1 - p_{23}} \left(p_1 T^{(1)}+2 W^{(1)} x\right)}{p_{12} p_{23}}\\
 \\
\dps \frac{ x^{1 + p_{13}} \left(p_2 T^{(1)}+2 W^{(1)} x\right)}{p_{13} p_{23}}\;\;& \dps \frac{ x^{1 + p_{23}} \left(p_1 T^{(1)}+2 W^{(1)} x\right)}{p_{13} p_{23}}\;\;& \dps \frac{ 2 x\left(  x W^{(1)} - p_3 T^{(1)}\right)}{ p_{13} p_{23}}
\end{pmatrix} \,. 
\ee

\vspace{3mm}

The monodromy matrix (up to the first order) of the solutions $\psi_{i} (x,z) = \psi^{(0)}_{i} (x)+ \psi^{(1)}_{i} (x,z)$ has the form
\be
\label{l3}
 M_{ij} (z) = M_{ij}^{(0)} + M_{ij}^{(1)} =\delta_{ij} + I_{ij}, \quad I_{ij} = \int_{\Gamma} dy ~ T_{ij}(y, z)\;.
\ee
Few comments are in order. First, the zeroth order solutions have no branch points inside the contour $\Gamma$, so $M_{ij}^{(0)} =\delta_{ij}$.  Second, the diagonal integrals $I_{ii}, i=1,2,3$ are 0, as well as $I_{12}, I_{21}$ due to \eqref{rc}. Third, the remaining $I_{ij}$ are linear in $c(z)$, for example
\be
\ba{c}
\dps \frac{I_{13}}{8 i \pi} = \frac{\epsilon_1 \left(\alpha_H^2 (z-1)^2 \left(z -z^{\alpha_H}\right)+\alpha_H (z-1) \left((3 z-2) z^{\alpha_H}+2 (1-2 z) z\right)+z \left((3-2 z) z^{\alpha_H}+(3 z-4) z\right)\right)}{3 (z^{\alpha_H}-2) (z^{\alpha_H}-1) (z-1)^2 z}\\
\\
\dps  -\frac{\epsilon_2 \left(z \left(3 \alpha_H (z-1)^2+z (8-3 z)-6\right)-z^{\alpha_H} \left(\alpha_H^2 (z-1)^2+\alpha_H ((7-3 z) z-4)+z (2 z-3)\right)\right)}{3 (\alpha_H-2) (\alpha_H-1) (z-1)^2 z} \\
\\
+ \dps \frac{c(z) \left(z^{\alpha_H}-(\alpha_H-1) z^2+(\alpha_H-2) z\right)}{(\alpha_H-2) (\alpha_H-1) (z-1)},
\ea
\ee
and $I_{31}, I_{32}, I_{23}$ can be written down using the following relations
\be
\label{ch}
I_{31} = - I_{13}\big|_{p_1 \leftrightarrow p_3}, \qquad I_{32} = I_{23}\big|_{p_2 \leftrightarrow p_3}, \qquad I_{32} = - \frac{z^{2 p_{23}} p_{12}}{p_{23}}I_{13}\big|_{p_2 \leftrightarrow p_3}. 
\ee
Finally,  comparing the eigenvalues of the matrices \eqref{l3} and \eqref{MTC} give us the monodromy equation 
\be
\label{M3}
I_{13} I_{32} + I_{23} I_{31} = - 16 \pi^2 \epsilon^2_p. 
\ee

\subsection{The monodromy equation for the 4-pt block with one non-degenerate, two level-1 and one level-2 operators}
\label{sec:HL2}
Here we use the HL approximation for the BPZ-type equation considered in Section \bref{Sec:BPZ22}. We assume that 2 level-1 operators at points $z_3=0$ and $z_4=\infty$ are heavy; the level-2, non-degenerate operators at points $z_1 = 1$ and $z_2 = z$  and the intermediate operator are light
\be
\epsilon_2, \epsilon_p \ll \epsilon_H, \qquad q_{1,2}\ll  q_H,
\ee
where we take into account that $\epsilon_1 \ll q_1$ \eqref{HLd2}.
The HL expansion has the form 
\be
  \tilde \psi(x,z) = \tilde \psi^{(0)}(x) +  \tilde \psi^{(1)}(x,z) +..., \qquad \tilde{c}(z) = \tilde{c}^{(1)} (z) + ....
\ee
In the zeroth order we get the following equation
\be
\label{N2F01}
\left(D^{(0)}\right)^2 \tilde \psi^{(0)}(x) =0,
\ee
while in the first order 
\be
\label{N2F11}
\left( D^{(0)} \right)^2 \tilde \psi^{(1)}(x,z) = - D^{(0)} \tilde{D}^{(1)} \tilde \psi^{(0)}(x) - \frac{2 q_1 q_H (x-1)^2 (z-2) (x-z)^2}{x^2 (z-1)^3} D^{(0)} \tilde \psi^{(0)}(x),
\ee
where the operator $D^{(0)}$ is given by \eqref{0o} and the operator $D^{(1)}$ is 
\be
 \tilde{D}^{(1)} = 4 \tilde{T}^{(1)} \frac{d}{dx} +  2 \frac{d\tilde{T}^{(1)}}{dx} - 4\tilde{W}^{(1)}, 
 \ee
where 
\be
\ba{c}
\dps \tilde{T}^{(1)} = \frac{(x-1) \left((z-1) z \tilde{c}(z) (z-x)+ \epsilon_2 \left(-2 x z+x+z^2\right)\right)}{(z-1)^2}, \quad \tilde{W}^{(1)} = \frac{1}{4} \left(\frac{q_2-2 q_1}{z-1}+\frac{q_1 x}{x-1}+\frac{q_2 x}{x-z}\right)  \\
\\
\dps +  \frac{\epsilon_2\left(x (2 z-1)-z^2\right) \left(4 \epsilon_H \left(5 x^2-3 x (z+1)+z\right)+3 q_H (x-1) (x-z)\right) }{8 \epsilon_H x (z-1)^2 (z-x)}  \\
\\
\dps + \frac{ z \tilde c(z) \left(4 \epsilon_H \left(5 x^2-3 x (z+1)+z\right)+3 q_H (x-1) (x-z)\right)}{8 \epsilon_H x (1-z)}. 
\ea 
\ee

Few comments are in order. First, the equation \eqref{N2F01} order has a factored form and is similar to the equation \eqref{0o}. Among the \eqref{N2F01} solutions
\be
\tilde \psi^{(0)}_i(x) = x^{1+p_i}, \quad \text{and} \quad \tilde \psi^{(0)}_j(x) = x^{1+p_j} \log x, \quad i, j = \overline{1,3}, 
\ee
where $p_i$ are given by \eqref{rc}, we choose only $ \tilde \psi^{(0)}_i(x)$, since their monodromy agrees with the given matrix \eqref{MTC} in the zeroth order in the HL approximation ($\epsilon_p =0$). Second, since for such solutions it is true that $D^{(0)} \tilde \psi^{(0)}(x)=0$, the equation \eqref{N2F11}
also factorizes into the form $D^{(0)} \left( D^{(0)} \tilde \psi^{(1)}(x,z) + \tilde{D}^{(1)} \tilde \psi^{(0)}(x) \right)=0$. Thus, we have only 3 branches of solutions that satisfy the monodromy properties
\be
\label{s2}
\tilde \psi^{(1)}_{i} (x,z) = \tilde \psi_{j}^{(0)}(x) \int \tilde{T}_{ij}(y, z) \; dy\;,
\ee
where $\tilde{T}_{ij}(y, z)$ has the same form as \eqref{mf1} with a replacement $T^{(1)} \rightarrow \tilde T^{(1)}, W^{(1)} \rightarrow \tilde  W^{(1)}$. 
Having this, one can construct the monodromy matrix up to the first level and the monodromy equations, which have the form 
 \be
 M_{ij} =  M^{(0)}_{ij} +  M^{(1)}_{ij} = \delta_{ij} + \tilde I_{ij}, \quad  \tilde I_{ij} = \int \tilde{T}_{ij}(y, z) \; dy. 
 \ee
Explicit expressions for $\tilde I_{ij}$ are massive, so we do not write down the explicit form of the monodromy equation. We consider the particular case in Section \bref{sec:two}.

\subsection{The classical $4$-pt blocks}
\label{Sec:MP}
\subsubsection{The  classical $4$-pt  blocks with level-1 operators}
For this case the monodromy equation \eqref{M3} is a quadratic equation for $c(z)$, which is similar to the Virasoro case \cite{Hijano:2015rla}. The equation can be solved exactly for any parameters $\epsilon_{2}, q_2$, $\epsilon_1$, $\epsilon_p$. In general, the solution for the accessory parameter is cumbersome and difficult to integrate.  

The relatively simple case occurs when the second operator is chosen to be level-1 semi-degenerate, i.e. $\epsilon_2 = \epsilon_1$, $q_2 = \epsilon_1/3$. For the case the accessory parameter has the form 
 \be
 \ba{c}
 \label{l1a}
 \dps c(z) = \frac{\epsilon_1}{3} \left(\frac{\alpha_H}{z}-\frac{3 \left(\alpha_H z^{\alpha_H-1}-2 (\alpha_H-1) z+\alpha_H-2\right)}{z^{\alpha_H}-(\alpha_H-1) z^2+(\alpha_H-2) z}\right) 
 \\
\dps  + \frac{\epsilon_p (\alpha_H^2-3 \alpha_H+2) (z-1) z^{\frac{\alpha_H-1}{2}}}{2 \sqrt{\alpha_H (z-1)-z+2} \left(z^{\alpha_H}-(\alpha_H-1) z^2+(\alpha_H-2) z\right)}. 
 \ea
 \ee
The classical block $f(z|\epsilon_1, \epsilon_p)$ is found by integrating the accessory parameter above
\be
\label{r1}
f(z|\epsilon_1, \epsilon_p) = \epsilon_1 \log \left( \frac{z^{\alpha_H/3}}{z^{\alpha_H}-(\alpha_H-1) z^2+(\alpha_H-2) z}\right) - \frac{\epsilon_p}{2} \log \frac{1 +z^{\frac{1-\alpha_H}{2}} \sqrt{\alpha_H (z-1)-z+2}}{1 - z^{\frac{1-\alpha_H}{2}} \sqrt{\alpha_H (z-1)-z+2}}. 
\ee
For the case $\epsilon_p=0$ this function coincides with the identity $4$-pt block, derived in \cite{deBoer:2014sna}. 

\subsubsection{The  classical $4$-pt blocks with one non-degenerate, one level-2 and 2 level-1 operators}
\label{sec:two}
For this case the monodromy equation is a third-order equations, so we limit ourselves to the case of $\epsilon_p=0$. The result for the accessory parameter associated with the $4$-pt classical block, considered in Section \bref{sec:HL2}, is
\be
\ba{c}
\dps \tilde c(z) = \frac{\alpha_H \left(\epsilon_2-q_1\right)}{z} + \frac{\left(\epsilon_2-3 q_1\right) (\alpha_H-\alpha_H^2+2) z^{\alpha_H+1}+(\alpha_H-1) \alpha_H z^{\alpha_H}-2 z^2)}{2((\alpha_H (z-1)-2 z+1) z^{\alpha_H+1}+z^3)} \\
\\\dps -\frac{\left(\epsilon_2+3 q_1\right) \left(\alpha_H z^{\alpha_H-1}-2 (\alpha_H-1) z+\alpha_H-2\right)}{2(z^{\alpha_H}-(\alpha_H-1) z^2+(\alpha_H-2) z)}. 
\ea
\ee
The $4$-pt classical block reads 
\be
\ba{c}
\label{r2}
\dps \tilde{f}(z|\epsilon_2, q_1, q_2) = \frac{\alpha_H \left(\epsilon_2-q_1\right)}{2} \log (z) - (\epsilon_2+3 q_1) \log \left(z^{\alpha_H}-(\alpha_H-1) z^2+(\alpha_H-2) z\right)
\\
-\left(\epsilon_2-3 q_1\right) \log \left((\alpha_H (z-1)-2 z+1) z^{\alpha_H}+z^2\right), 
\ea
\ee
which is consistent with the identity $4$-pt block, derived in \cite{deBoer:2014sna}. 
\vspace{-3mm}
\section{Concluding remarks}
\label{sec:MP}

In this paper, we investigated 4-point $\mathcal{W}_3$ classical blocks, focusing on two cases: 1) blocks with three level-1 semi-degenerate operators and 2) blocks with two level-1 and one level-2 semi-degenerate operators. Calculating such blocks in the classical limit requires constructing BPZ-type equations for auxiliary $5$-pt blocks with the fully degenerate operator. The equations involve additional blocks with $\mathcal{W}_3$ descendants operators. In Section \bref{sec:BPZ}, we explicitly argued that these blocks have the exponential form \eqref{CL}. This allowed us to rewrite the BPZ-type equations as differential equations: a third-order equation \eqref{BPZ} for the first case, and a sixth-order equation for the second. 

In Section \bref{sec:HL}, we solved the BPZ-type equations using HL perturbation theory. For both cases, we considered two heavy level-1 semi-degenerate operators, treating the remaining operators as light. We solved the equations perturbatively (to first order). The monodromy properties of these solutions determine the accessory parameters of the blocks. By integrating these accessory parameters, we obtained explicit expressions for non-identity blocks, which complement the results in \cite{deBoer:2014sna}. 

The formula \eqref{r1} for the classical block with level-1 semi-degenerate operators generalizes the results in \cite{deBoer:2014sna} by implementing the case with non-identity intermediate operators. A natural extension of this work is to consider its implications within the framework of higher-spin holography \cite{Campoleoni:2010zq, Campoleoni:2011hg, Gaberdiel:2012yb}. While the identity block has been rigorously shown to be dual to a Wilson line in AdS \cite{deBoer:2014sna}, the status of non-identity blocks in the context of AdS/CFT remains an open question. It would also be interesting to derive the classical block \eqref{r2} by applying first the classical and then the HL approximation to the corresponding quantum 4-point $\mathcal{W}_3$ block with one level-2 semi-degenerate operator, studied in \cite{Belavin:2016wlo} and to analyze this case in the context of AdS/CFT. 

We note that there exists a relation between $\mathcal{W}_N$ conformal blocks and topological string partition functions. Analysis of conformal blocks from this perspective have been carried out in \cite{Coman:2017qgv, Coman:2019eex}. This representation is expected to be useful for studying the classical and heavy–light limits in the presence of level $2$ semi-degenerate operators. We leave a detailed investigation of this direction for future work.

\paragraph{Acknowledgments.} M.P. work was supported by the Russian Science Foundation (RSF) grant № 25-72-10177. We thank K. Alkalaev and E. Pomoni for drawing our attention to related developments regarding the exponentiation of conformal blocks and the connections between general $\mathcal{W}_N$ blocks and topological strings. M.P. would also like to thank the organizers of the conference "Fields and Strings 2025" for the hospitality and productive atmosphere. 

\providecommand{\href}[2]{#2}\begingroup\raggedright\endgroup

\end{document}